\documentclass[showpacs,twocolumn,preprintnumbers,amsmath,amssymb]{revtex4}
\usepackage{graphicx} 

\begin{document}

\title{\bf Screening effects in flow through rough channels}

\author{J. S. Andrade Jr.$^{1,2,3}$, A. D. Ara\'ujo$^{1}$, M. Filoche$^{2,3}$,
and B. Sapoval$^{2,3}$}

\affiliation{$^1$Departamento de F\'\i sica, Universidade Federal 
do Cear\'a, 60451-970 Fortaleza, Cear\'a, Brazil\\
$^2$Centre de Math\'ematiques et de leurs Applications,\\
Ecole Normale Sup\'erieure de Cachan, 94235 Cachan, France\\
$^3$Laboratoire de Physique de la Mati\`ere Condens\'ee, Ecole Polytechnique,\\
91128 Palaiseau, France}

\date{\today}

\begin{abstract}
A surprising similarity is found between the distribution of
hydrodynamic stress on the wall of an irregular channel and the
distribution of flux from a purely Laplacian field on the same
geometry. This finding is a direct outcome from numerical simulations
of the Navier-Stokes equations for flow at low Reynolds numbers in
two-dimensional channels with rough walls presenting either
deterministic or random self-similar geometries. For high Reynolds
numbers, when inertial effects become relevant, the distribution of
wall stresses on deterministic and random fractal rough channels
becomes substantially dependent on the microscopic details of the
walls geometry. In addition, we find that, while the permeability of
the random channel follows the usual decrease with Reynolds, our
results indicate an unexpected permeability increase for the
deterministic case, i.e., ``the rougher the better''. We show that
this complex behavior is closely related with the presence and
relative intensity of recirculation zones in the reentrant regions 
of the rough channel. 
\end{abstract}


\maketitle
Partial differential equations are basic in the mathematical
formulation of physical problems. The Laplace equation, for example,
is known for its relevance in many fields, namely electrostatics, heat
transport, heterogeneous catalysis, and electrochemistry. Although
rather simple in form, the Laplace equation can have highly
non-trivial solutions, specially if the boundary of the system
represents the surface of an irregular object. The role of this
particular complexity has been a theme of constant research interest
with recent important developments \cite{Sapoval94,Makarov85,Meakin85}.  
Much more complex, however, are the Navier-Stokes equations for the
description of hydrodynamic flow in irregular geometries. The aim of
the present work is first to reveal a surprising analogy between the
properties of the solutions of 2d Laplace and Navier-Stokes equations
when flow at low Reynolds and proper boundary conditions are imposed
on the {\it same geometry}. In a second step, we present results
obtained at higher Reynolds numbers and for distinct types of surface
geometry.

\begin{figure}
\includegraphics[width=8cm]{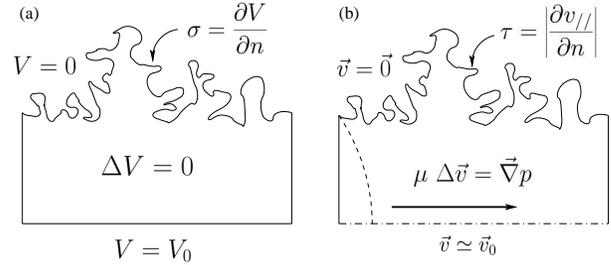}
\caption{The two different problems with similar solutions. In 
(a) we show a capacitor with an irregular electrode. The local charge
is obtained from the numerical solution of the Laplace equation with
Dirichlet boundary conditions. In (b) we see the analogous problem of
flow at low Reynolds number. The stress parallel to the wall channel
can be calculated from the solution of the continuity and Stokes
($Re=0$) equations.}
\end{figure}

The situations we compare and find to be quantitatively similar are
displayed in Fig.~1. Figure~1a pictures the simplest Laplacian
problem, namely that of a capacitor with a pre-fractal electrode. The
complex feature here is the distribution of charge on the irregular
electrode or, in mathematical terms, the distribution of the harmonic
measure. The recent research on this field has been mainly dedicated
to the application of Laplacian transport towards and across irregular
interfaces \cite{Sapoval94,Andrade01,Filoche00}. The system depicted
in Fig.~1b corresponds to Stokes flow in a symmetric rough channel
with the same geometry as in Fig.~1a. The hydrodynamic quantity which
is found to be distributed similarly to the harmonic measure in
Fig.~1a is the viscous stress along the channel boundary.

To obtain the potential field for the problem in Fig.~1a, one must
compute the solution of Laplace equation with a potential $V=1$ on the
counter electrode and zero potential on the irregular electrode
(Dirichlet boundary condition). From that, it is then possible to
calculate the charge $\sigma_{L}^{i}$ on each elementary unit $i$ of
the wall and its corresponding normalized counterpart, $\phi_{L}^{i}
\equiv \sigma_{L}^{i}/\sum \sigma_{L}^{j}$. As previously shown
\cite{Evertsz92}, the distribution of $\phi_{L}$ along the irregular
electrode is strongly nonuniform as a consequence of screening
effects, being in fact characterized as multifractal. This is a 
typical situation where the deep regions of the irregular surface
support only a very small fraction of the total charge, as opposed 
to the more exposed parts.

The hydrodynamic question posed here has been triggered by the idea
that the design of flowing systems should include the influence of the
surface geometry as a possibility for optimal performance. For this,
we investigate the flow in a duct of length $L$ and width $h$ whose
delimiting walls are identical pre-fractal interfaces with the
geometry of a square Koch curve (SKC) \cite{Evertsz92}. The
mathematical description for the fluid mechanics in this channel is
based on the assumptions that we have a continuum, Newtonian and
incompressible fluid flowing under steady state conditions. The
relevant physical properties of the fluid are the density $\rho$ and
the viscosity $\mu$. In our simulations, we consider non-slip boundary
conditions at the entire solid-fluid interface. In addition,
gradientless boundary conditions are assumed at the exit $x=L$,
whereas a parabolic velocity profile is imposed at the inlet of the
channel. The numerical solution for the velocity and pressure fields
in the rough channel is obtained through discretization by means of
the control volume finite-difference technique \cite{Patankar80}. For
our complex geometry, this problem is solved using a structured mesh
based on quadrangular grid elements. For example, in the case of the
channel with walls that are fourth-generation SKC curves, a mesh of
approximately a million elements adapted to the geometry of the
interface generates satisfactory results when compared with numerical
meshes of higher resolution.

\begin{figure}
\includegraphics[width=8cm]{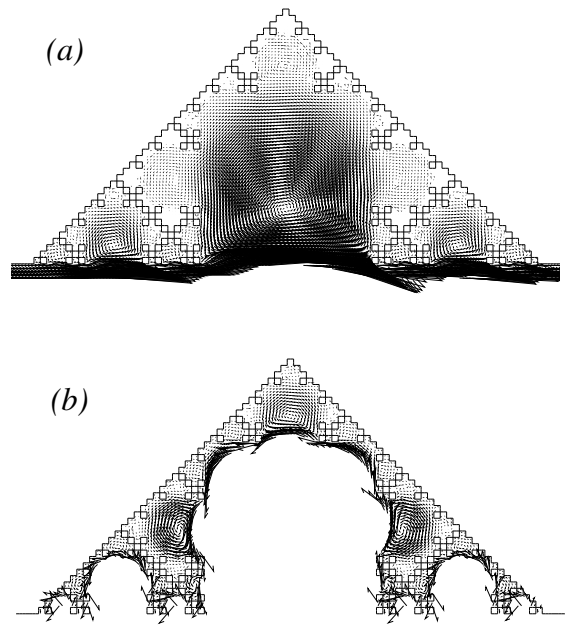}
\caption{(a) Vortices in the reentrant zones of the upper half of the 
(symmetric) pre-fractal roughness channel. The fluid flows steadily
from left to right at a low Reynolds number, $Re=0.01$. (b) Sequence
of the smaller eddies at the details of the roughness wall shown in
(a). These structures can only be viewed by rescaling the magnitude of
the velocity vectors because their intensities fall off in geometric
progression.} 
\end{figure}

The Reynolds number is defined here as $Re\equiv{\rho V h /
\mu}$, where $V$ is the average velocity at the inlet. In Fig.~2a 
we show the velocity vector field at low $Re$ located at the
self-similar reentrant zones that constitutes the roughness of the
irregular channel. Indeed, as depicted in Fig.~2b, by rescaling the
magnitude of the velocity vectors at the details of the roughness
wall, we can observe fluid layers in the form of consecutive
eddies. Although much less intense than the mainstream flow, these
recirculating structures are located deeper in the system, and
therefore experience closer the landscape of the solid-fluid
interface. More precisely, viscous momentum is transmitted laterally
from the mainstream flow and across successive laminae of fluid to
induce vortices inside the fractal cavity. These vortices will then
generate other vortices of smaller sizes whose intensities fall off in
geometric progression \cite{Moffat64,Leneweit99}.

Once the velocity and pressure fields are obtained for the flow in the
rough channel, we can compute the normalized stress $\phi_{S}^{i}$ at
each elementary unit $i$ of the wall, $\phi_{S}^{i} \equiv
\tau_{S}^{i}/\sum \tau_{S}^{j}$, where the sum is over the total 
number $L_{p}$ of perimeter elements, the magnitude of the local
stress is given by $\tau_{S}= |\partial v_{\|}/\partial n|$, the
derivative is calculated at the wall element, $v_{\|}$ is the local
component of the velocity that is parallel to the wall element, and
$n$ is the local normal coordinate. The semi-log plot in Fig.~3a shows
that the spatial distribution of normalized stresses at the interface
is highly heterogeneous, with numerical values in a range that covers
more than five orders of magnitude. Also shown in Fig.~3a is the
variation along the interface of the normalized Laplacian fluxes
$\phi_{L}$ crossing the wall elements of a Laplacian cell with
Dirichlet boundary conditions. The astonishing similarity between
these two distributions clearly suggests that the screening effect in
flow could be reminiscent of the behavior of purely Laplacian
systems. As shown in Fig.~3b, this analogy is numerically confirmed
through the very strong correlation between local stresses and
Laplacian fluxes. These measures follow an approximately linear
relationship, namely $\phi_{S} \propto \phi_{L}$.

\begin{figure}
\includegraphics[width=7cm]{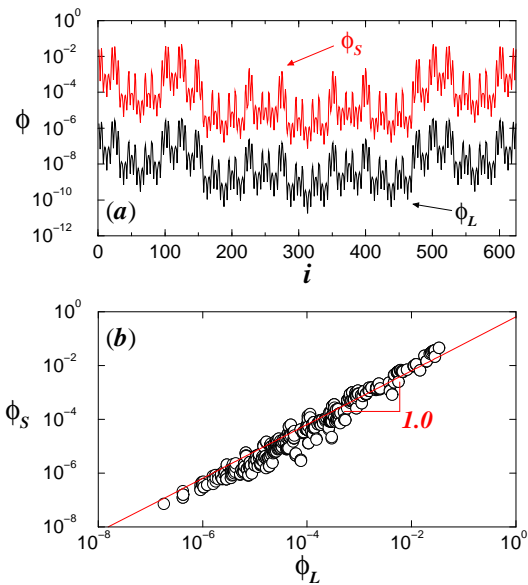}
\caption{(a) The (dark) black line is the curvilinear distribution 
of the logarithm of the normalized shear stresses $\phi_{S}$ on the
interface along one of the two (symmetric) square Koch curves
corresponding to the channel walls. The (light) red line gives the
distribution of the logarithm of the normalized Laplacian charges
$\phi_{L}$ for the analogous electrostatic problem. For better
visualization, the distribution of $\phi_{L}$ has been shifted
downwards. (b) Double-logarithmic plot of $\phi_{S}$ versus
$\phi_{L}$, with the red line indicating their linear relationship.}
\end{figure}

A further analogy can be drawn from the discussion of the notion of
{\it active zone}, as the zone which supports the majority of the
charge or current (for the Laplacian field) as well as the shear
stress here \cite{Sapoval94}. For two-dimensional systems subjected to
Dirichlet's boundary condition, a fundamental step towards the
understanding of the purely Laplacian problem is the mathematical
theorem given by Makarov \cite{Makarov85}. This theorem states that
{\it the information dimension of the harmonic measure on a singly
connected interface of arbitrary geometry in $D=2$ is exactly equal to
1}. In practical terms, it essentially says that, whatever the shape
(perimeter) of an interface, the size of the region where most of the
activity takes place is of the order of the overall size $L$ of the
system. Here we define an {\it active length} $L_{a}$ as,
\begin{equation}
L_{a}\equiv 1/\sum_{i=1}^{L_{p}}(\phi_{S}^{i})^2~~~~
(1 \leq L_{a} \leq L_{p})~.
\label{active_length}
\end{equation}
If $L_{a}$ is equal to the wall perimeter $L_{p}$, the entire wall
works uniformly. The theorem of Makarov indicates that, on the
opposite, for a purely Laplacian field, $L_{a} \approx L$. The active
length therefore provides an useful index to quantify the interplay
between the flow and the complex geometry of the interface at the
local scale. The results in Fig.~4a (open circles) show that the value
of $L_{a}$ for the square Koch curve remains approximately constant at
$L_{a}/L=0.55$ for low and moderate Reynolds numbers. This value is
consistent with our screening analogy because it indicates that the
hydrodynamic stress is mainly concentrated in a subset of the wall
whose size is of the order of the system size $L$. Only at higher $Re$
values, when inertial forces become comparable to viscous forces, one
can observe a small increase in $L_{a}$. The stress becomes slightly
less localized due to the higher relative intensities of the vortices
inside the deeper reentrant zones, when compared with the intensities
of the correspondent flow structures at low Reynolds conditions.

\begin{figure}
\includegraphics[width=8cm]{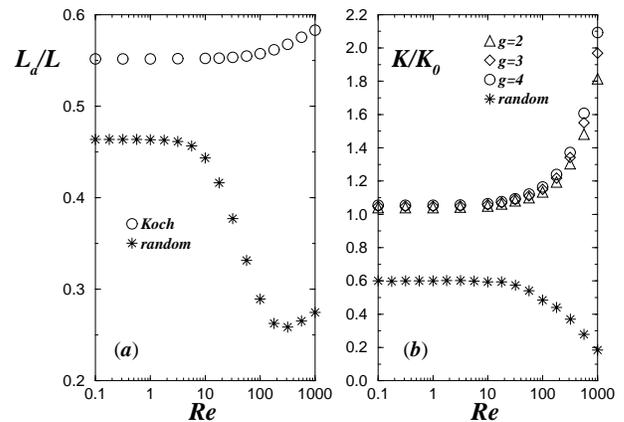}
\caption{(a) Semi-log plot showing the dependence of the active length 
$L_{a}$ on the Reynolds number $Re$ for the square Koch channel (empty
circles) and random Koch channel (stars). In (b) we show the semi-log
plot of the variation of the normalized permeability with $Re$ for the
random Koch curve (stars) and the second (triangles), third (diamonds)
and fourth (circles) generations of the square Koch channel. One observes
that the higher is the generation of the pre-fractal SKC, the larger is
its permeability.}
\end{figure}

The usual approach to describe single-phase fluid flow in irregular
media (e.g., porous materials and fractures) is to characterize the
system in terms of a macroscopic index, namely the permeability $K$,
which relates the average fluid velocity $V$ with the pressure drop
$\Delta P$ measured across the system, $V=-K \Delta P/\mu L$.
Figure~4b (open circles) shows that the permeability of the rough
channel for low and moderate $Re$ remains essentially constant at a
value that is slightly above but very close to the reference
permeability of a two-dimensional smooth channel, namely ($K/K_{0}
\simeq 1$), with $K_{0} \equiv h^{2}/12$ \cite{Bird60}. 
Above a transition point at about $Re \simeq 10$, the change in
permeability reflects the onset of convective effects in the flow, and
therefore the sensitivity of the system to these inertial
nonlinearities. Surprisingly, one observes that, instead of decreasing
with $Re$ (i.e., a behavior that is typical of disordered porous
media), the permeability of the SKC substantially increases. Moreover,
as shown in Fig.~4b, the higher the generation of the SKC, the higher
is the permeability of the channel for a fixed value of $Re$ above the
transition. In other words, ``the rougher the better''. These results
show that the screening effect of the hierarchical SKC geometry on the
flow can be understood in terms of a reduction in the effective
non-slip solid-fluid interface. In other words, we can imagine that
each vortex present in a given generation of the SKC is in fact
replacing one or a set of highly dissipative (non-slip) wall elements
of previous SKC of lower generations.

Next we study the fluid flow through a rough channel whose walls are
composed of 10 successive and distinct realizations of the random Koch
curve (RKC) of third generation. As in the previous case with the
deterministic SKC walls, this task is performed here through
finite-differences \cite{Patankar80}, but now with an unstructured
mesh of triangular grid elements based on a Delaunay
network. Interestingly, the results shown in Fig.~4a (full circles)
indicate that the active length $L_{a}$ of the wall stress calculated
for the entire irregular interface geometry at low values of $Re$,
$L_{a}/L \simeq 0.46$, is not substantially different from the SKC
case, where $L_{a}/L=0.55$. In Fig.~5 we show that the ratio $L_{a}/L$
calculated individually for each of the 10 wall subsets composing the
rough channel does not vary significantly from one unit to
another. This is a rather unexpected behavior, specially if we
consider the complexity of the different geometries involved (see
Fig.~5). A similar effect has been observed for the purely Laplacian
problem in a random geometry \cite{Filoche00}. These facts strongly
reinforce our conjecture that the rationale behind the screening
effect in flow through rough channels should be based on a general
conceptualization that is quite similar to the well established theory
for Laplacian systems \cite{Makarov85}.

\begin{figure}
\includegraphics[width=8cm]{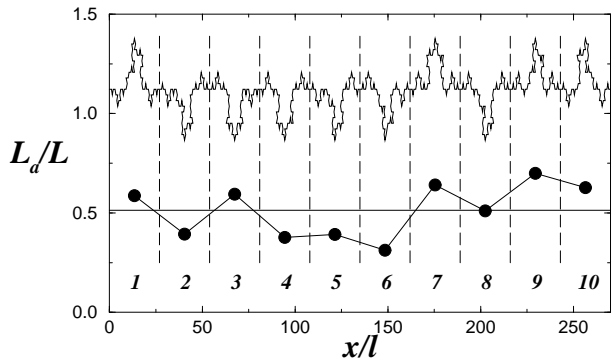}
\caption{The active length $L_{a}$ for each of the 10 wall subsets 
composing the entire geometry of the random Koch curve channel shown
at the top. The dashed lines indicate the boundaries between
consecutive subsets, as numbered below. The fluid flows from left to
right at $Re=0.1$.}
\end{figure}

As depicted in Fig.~4a, by increasing the Reynolds number, the
departure from Stokes flow due to convection at $Re \simeq 10$ results
initially in the decrease of $L_{a}/L$ (calculated over the entire
surface) down to a minimum of approximately $0.25$ at $Re \simeq
200$. This behavior indicates the presence of long-range flow
correlations imposed by inertia among successive wall subsets. More
specifically, due to the randomness of these interface units, we can
observe either ``inward'' (e.g., the wall subsets 2, 3, 4, 5, 6 and 8
in Fig.~4) or ``outward'' protuberances (e.g., the wall subsets 1, 7,
9 and 10 in Fig.~4) composing the roughness of the channel. Due to the
symmetry of the system, the inward elements generate bottlenecks for
flow.  At high values of $Re$, the effect of inertia is to induce {\it
flow separation lines} between the mainstream flow at the center of
the channel and the flow near the wall, that can be as large as the
largest distance between two consecutive bottlenecks. These
exceedingly large ``stagnation regions'' are responsible for the
initial decrease in the active length. If we increase even more the
Reynolds number, the relative intensities of the vortices in these
regions starts to increase. As in the case of the SKC channel,
$L_{a}/L$ starts to increase due to a more distribution of shear
stress near the wall.

Finally, in Fig.~4b we show the variation with $Re$ of the
permeability for the RKC channel (full circles). In this case, we also
observe a transition from linear (constant $K$) to nonlinear behavior
that is typical of experiments with flow through real porous media and
fractures \cite{Dullien79}. Contrary to the results obtained for the
SKC channel, however, the value of $K$ calculated for low $Re$ values
is significantly different and smaller than the reference value
$K_{0}$ of the corresponding smooth channel ($K/K_{0} \simeq
0.6$). Once more, this is a consequence of the presence of several
bottlenecks in the channel, which drastically reduce the effective
space for flow. At high Reynolds, this difference is amplified due to
inertial effects.

In summary, we have investigated the effect of deterministic as well
as random roughness of 2D channels on local as well as macroscopic
flow properties. Our main results are threefold. First, at low
Reynolds numbers there exists a very close analogy between the spatial
distribution of the local stress on the rough walls and the
distribution of charge resulting from the solutions of the Laplace
equation for the electric potential in the same geometry. Second, for
a fractal deterministic roughness, a surprising increase of the
permeability with Reynolds is observed. Moreover, this effect is
augmented by increasing the fractal generation of the channel wall so
that, paradoxically, ``the rougher the better''. Such results could
find practical applications in microfluidics and help, for example, to
understand the enhanced hydrodynamical features underlying shark-skin
effects.
 
We thank the Brazilian agencies CNPq, CAPES, FUNCAP and FINEP for
financial support.

\end{document}